# Demonstration of mode splitting in an optical microcavity in aqueous environment

Woosung Kim[1], Sahin Kaya Ozdemir[2], Jiangang Zhu, Lina He, and Lan Yang[3]

Department of Electrical and Systems Engineering, Washington University, St. Louis, Missouri 63130, USA

1. wk1@cec.wustl.edu
2. ozdemir@ese.wustl.edu
3. yang@ese.wustl.edu

**Abstract** Scatterer induced modal coupling and the consequent mode splitting in a whispering gallery mode resonator is demonstrated in aqueous environment. The rate of change in splitting as particles enter the resonator mode volume strongly depends on the concentration of particle solution: The higher is the concentration, the higher is the rate of change. Polystrene nanoparticles of radius $50\,nm$ with concentration as low as $5 \times 10^{-6}\,wt\%$ have been detected using the mode splitting spectra. Observation of mode splitting in water paves the way for constructing advanced resonator based sensors for measuring nanoparticles and biomolecules in various environments.

In a whispering gallery mode (WGM) optical resonator, light circulates along the curved boundary between the resonator material and the surrounding medium by near-ideal total internal reflection. At each revolution the light propagating inside the resonator interacts with the surrounding scatterers or dipoles. The ultrahigh quality factor ($Q$) and small mode volume ($V$) of WGM resonators provide enhanced light-matter interaction making the detection of even single binding events possible [1]. Optical WGM resonators thus have emerged as a promising sensing technology and have recently been under intensive investigation [2].

The detection method, based on measuring shift of resonance wavelength of a single WGM, has been largely used to recognize biomolecules and subwavelength particles directly adsorbed on the resonator surface. Arnold *et al.* [3] demonstrated that the resonance wavelength of a WGM in a microsphere shifts in response to the excess polarizability induced by a molecule entering the resonator mode volume and discussed the possibility of single-protein detection using this reactive mechanism. Following this work, Armani *et al.* [4] observed using a microtoroid single molecule sensitivity attributed to a thermo-optic effect. A recent theoretical work by Arnold and co-workers [5] have shown that under the experimental conditions employed the thermo-optic effect results in a smaller shift than the reactive effect, and a new mechanism is necessary to explain the observations. Vollmer *et al.* [6] demonstrated single virus detection using a microsphere. Meanwhile, WGM resonators of various shapes, such as microring [7], microdisk, microsphere [8] and micro-capillary [9,10], have been used to detect a wide range of large biomolecules, biomarkers, bacteria and viral particles.

Mode splitting in high-$Q$ WGM resonators has emerged as an alternative to resonance-frequency-shift method and has been demonstrated to yield label-free and highly sensitive detection of particles with radii as small as $30\ nm$ with single particle resolution [11]. Mode splitting occurs due to the coupling of counter-propagating doubly degenerate WGMs via the scattering of light from a sub-wavelength scatterer entering the resonator mode volume [12-15]. This modal coupling lifts the degeneracy and creates two standing wave modes whose resonance frequencies and linewidths differ by $2|g| = -\alpha f^2(r)\omega/V$ and $2\Gamma_R = -\alpha^2 f^2(r)\omega^4/(3\pi v^3 V)$, respectively. Here, the polarizability $\alpha$ is defined as $\alpha = 4\pi R^3\ (n^2 - n_e^2)\ /\ (n^2 + 2n_e^2)$, for a single particle of radius $R$ and refractive index $n$ in the surrounding medium of refractive index $n_e$, $f^2(r)$ is the normalized mode distribution, $v$ is the speed of light in the medium, and $\omega = 2\pi c/\lambda$ is the angular frequency of the resonant, $\lambda$ and c being the wavelength of WGM before splitting and the speed of light in vacuum, respectively [11,16,17]. Polarizability of a scatterer is calculated as $\alpha = -\left(\frac{\Gamma}{g}\right)\left(\frac{\lambda}{n_e}\right)^3 (3/8\pi^2)$ from which one can estimate the size. Advantages of mode splitting method over the resonance-frequency-shift method are the accurate estimation of the size regardless of the location of the particle in the resonator mode volume and the robustness of the mode splitting spectra against interfering perturbations (e.g., laser and detector noises, temperature fluctuations which uniformly affect the resonator) [18].

Up to date, all the experiments on mode splitting have been performed in air where it is easier to satisfy the mode splitting resolvability criterion $2|g| > \Gamma + (\omega/Q)$. Demonstration of mode splitting in other media such as aquatic environment would open new possibilities for diverse applications, such as bio-chemical and bio-molecular sensing, detection and characterization of nanoparticles in liquid solutions. In this Letter, we report the demonstration of mode splitting in aquatic environment using a high-Q microtoroidal resonator and investigate the effect of particle concentration on mode splitting spectrum.

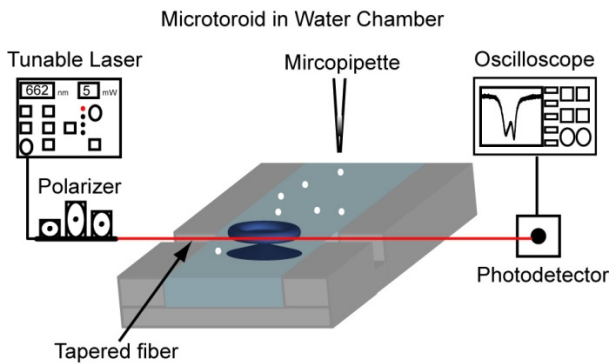

Figure 1 Schematics of the experimental setup. Micro-pipette is used to inject nanoparticles into the microaquarium.

The WGM optical microresonators used in our experiments are silica microtoroids (Diameter: 80-100 μm) fabricated using a three-step process: (i) photolitography to form circular silica pads over silicon wafer; (ii) selective isotropic etching of silicon with xenon difluoride gas to form undercut silica microdisks; and (iii) $CO_2$ laser reflow to turn microdisks into microtoroid [19]. A schematic diagram of our experimental setup is shown in Fig. 1. Light from a tunable laser $\lambda = 670nm$ is coupled into and out of the microtoroid using a fiber taper. The gap between the resonator and the fiber taper is controlled by a 3D nanopositioning system. A photodetector connected to an oscilloscope is used to monitor the transmission spectra while the wavelength of the laser is scanned. After characterizing the microtoroid in air, we immersed the chip and the fiber taper into a microaquarium filled with water, and placed a cover slip on the top. The power before and after the the taper section was measured as $\sim 0.6 mW$ and $\sim 0.4 mW$, respectively. The taper-resonator gap is adjusted to minimize thermal effects. The power coupled into the resonator is estimated as $\sim 0.18 mW$ which is sufficient to exert carousel forces for the binding of particles to the resonator [20]. After the characterization of the resonator in water, solution containing polystrene nanoparticles (PS: refractive index $n = 1.59$ and average size $R = 50 nm$ was injected into the microaquarium. The evanescent tail of the resonant WGM probes the water environment and interacts with the particles once they enter the mode volume or bind onto the toroidal surface. This then leads to modal coupling which shows itself as a doublet (split modes) in the transmission spectra. Figure 2(a) and (b) show typical transmission spectra obtained for a microtoroid in air and in water, respectively, before the nanoparticle solution was injected. The higher loss in water shows itself as a 10-fold decrease in the $Q$ of the resonator, i.e., the quality factor reduced from $Q_{air} \sim 7 \times 10^6$ in air to $Q_{water} \sim 7 \times 10^5$ in water. Upon the arrival of a PS particle into the mode volume, the single resonant mode observed in the transmission spectra undergoes mode splitting which leads to a doublet. The splitting spectrum shown in Fig.2c has $2|g| = 440.9 \, MHz$ and $\Gamma = |\gamma_1 - \gamma_2| = 1.9 \, MHz$. For a single $R = 50 nm$ PS particle, we theoretically estimate $2|g| = 12.3 \, MHz$ assuming that the particle resides on the mode maximum, $f(r) = 0.36$. This is much smaller than the measured value in Fig.2(c) suggesting multiple particle binding. Based on a multiple scatterer model [17,21], we estimate that at least $N = 25$ particles are needed to lead to a splitting of $2|g| = 440.9\ MHz$.

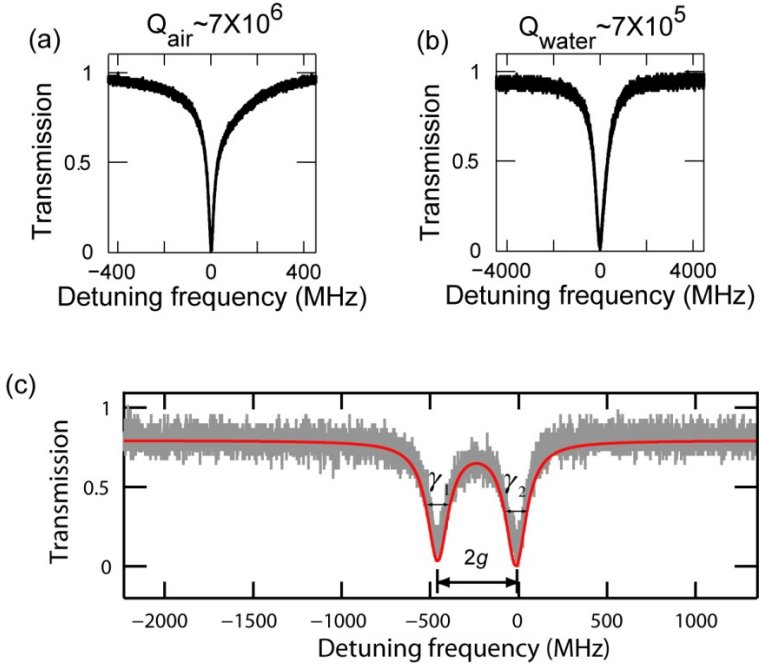

**Figure 2** Transmission spectra obtained for a microtoroid during the measurements in (a) air and (b) water before nanoparticle injection, and (c) in water after nanoparticle injection. The doublet in (c) is the manifestation of mode splitting.

Next, we test the response of the system to various concentrations of PS particles. Figure 3(a) and (b) show a series of transmission spectra obtained for concentrations of $5 \times 10^{-3}\ wt\%$ and $5 \times 10^{-6}\ wt\%$, respectively. Before the injection of the particle solution, there is no observable mode splitting. After particle injection, it takes a while for the particles to diffuse into the region closer to the resonator. As soon as a particle enters the resonator mode volume, a doublet is formed in the transmission spectra. As more particles bind, the amount of mode splitting $2g$ and the additional linewidth broadening $\Gamma$ change. Our theoretical studies and experiments in air on multiparticle binding have shown that at each single particle event $2g$ and $\Gamma$ undergo discrete jumps to higher or lower values depending on the location of the particle on the mode volume. This allows to resolve single binding events. However, we could not observe such discrete jumps in these experiments. Instead, mode splitting and linewidths increased continuously until there was no observable change in mode splitting or until mode splitting became unresolvable due to increasing dissipation as particle binding continued. Using the multiple scatterer model [17,21], we estimate that at least N=15 (N=60) and N=4 (N=29) particles are needed to observe the amount of mode splitting shown in the first (last) spectra of Fig. 3(a) and (b), respectively.

Figure 4 depicts the time-evolution of the amount of mode splitting $2g$ for different solution concentrations. Data acquisition was performed until mode splitting becomes unresolved. It is clearly seen that when the concentration is increased, the slope which quantifies the average binding rate increases too. At very high concentration ($10\ wt\%$), due to the larger number of particles in the microaquarium, binding rate is very high which leads to faster coverage of the resonator surface and shorter time for $2|g|$ to reach its maximum value. During this time $2|g|$ and $\Gamma$ continuously increase until a critical value of surface coverage (critical number of particles bind to the surface) is reached. Beyond this point, $2|g|$ suddenly starts decreasing while $\Gamma$ keeps on increasing. Thus, the split modes approach to each other with ever increasing linewidths and eventually merge together preventing our system to resolve them.

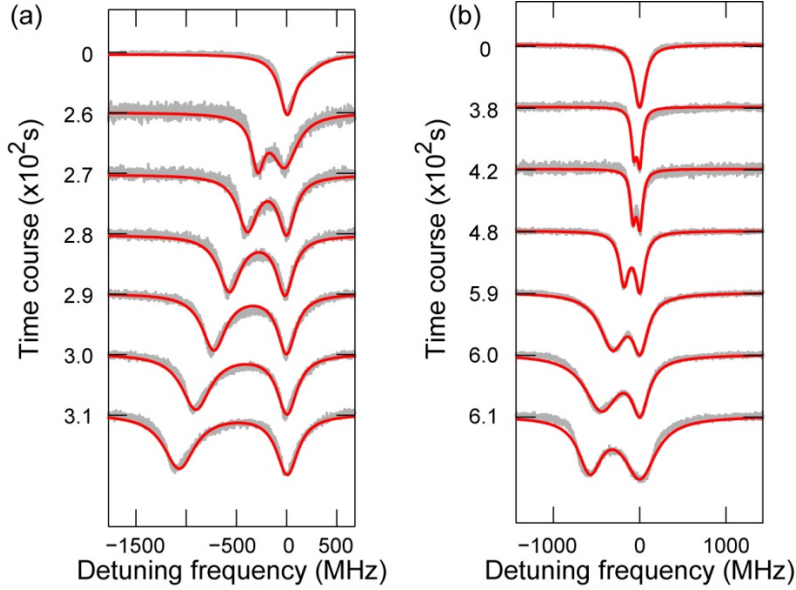

**Figure 3** A series of transmission spectra obtained for consecutive depositions of nanoparticles (PS, $R = 50nm$) into the resonator mode volume for particle concentrations of $5 \times 10^{-3}\ wt\%$ (left) and $5 \times 10^{-6}\ wt\%$ (right).

For lower concentrations ($5 \times 10^{-3}\ wt\%$, $6.5 \times 10^{-4}\ wt\%$, and $5 \times 10^{-6}\ wt\%$), on the other hand, binding rate and the number of particles expected to interact with the WGM field are smaller. We observe that at these low concentrations, the critical surface coverage is not reached during the measurement duration. Therefore, both $2|g|$ and $\Gamma$ increase on average; however the rate of increase of $\Gamma$ with the number of binding particles is so high that the linewidth of low frequency resonant mode undergoes continuous largening which eventually makes it impossible to resolve.

These observations in Fig.4 clearly show the concentration dependent binding phenomenon, and they coincide with the results of our numerical studies that on average $2|g| \propto \sqrt{N}$ and $\Gamma \propto N$. Thus, as the particles continuously enter the mode volume and bind to the resonator, $2|g|$ increases until a critical number of particles whereas $\Gamma$ always increases. As discussed also in Ref. [21], beyond this critical number mode splitting starts decreasing and finally becomes zero.

The probability of simultaneous binding of several particles onto the resonator surface is smaller in lower concentrations, i.e., the probability of one particle binding event is higher. Thus, one would expect to resolve single particle binding events for the resonators used in our experiments. Indeed, footprints of such events, although not as clear as one would expect, are seen in Fig. 4 for concentrations $6.5 \times 10^{-4}\ wt\%$, and $5 \times 10^{-6}\ wt\%$. When we look at the evolution of $2|g|$ much closer, we see that although on average there is a continuous increase, $2|g|$ shows small up and down jumps which is a consequence of the fact that $2|g|$ is dependent on the location of the particle on the mode volume, i.e., depending on the location of the incoming particles with respect to the position of the particles already on the resonator surface, $2|g|$ may decrease or increase. This observation suggests that with much faster data acquisition and much lower concentration, we could accomplish single particle resolution.

In conclusion, we have demonstrated scattering induced mode splitting in ultra-high $Q$ toroidal microcavities in aqueous environment. These results will open the way for detection of proteins, biomolecules and nanoparticles in aqueous environment or biomarkers in serum samples using mode-splitting phenomenon which provides a highly sensitive self-referencing method of detection. We believe that accomplishing these tasks at single particle resolution is within our reach with a faster data acquisition system and a more controlled fluid injection mechanism.

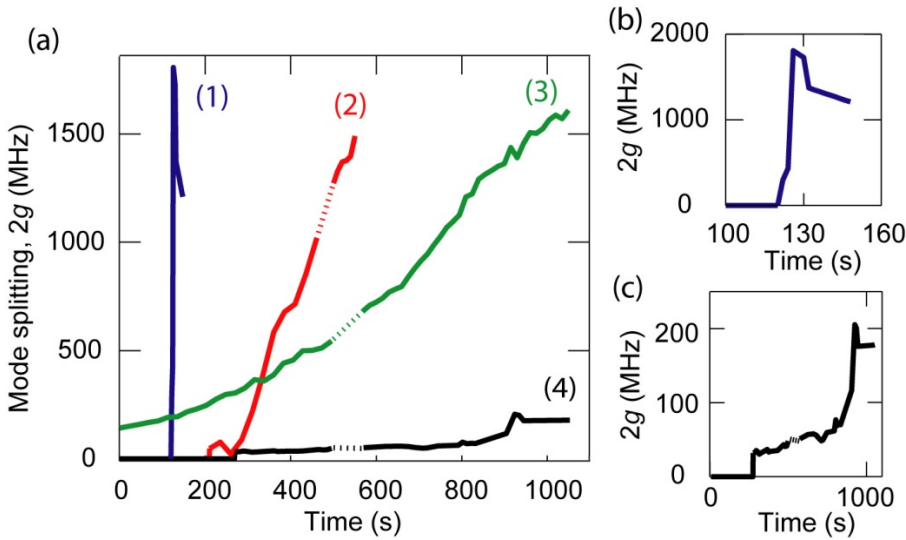

**Figure 4** Effect of concentration on the amount of mode splitting for PS particles of $R = 50nm$ at concentrations (1) $10\ wt\%$, (2) $5 \times 10^{-3}\ wt\%$, (3) $6.5 \times 10^{-4}\ wt\%$, and (4) $5 \times 10^{-6}\ wt\%$, in (a). Enlarged plots of the results in (1) and (4) are depicted, respectively, in (b) and (c). Non-zero $2|g|$ at zero-time (before the injection of PS particles) of (3) is due to intrinsic mode splitting which is caused by scattering centers formed due to surface inhomogeneity and contaminations.

The authors acknowledge the support from CMI and MISA at Washington University in St. Louis and NSF (Grant No. 0954941). This work was performed in part at the NRF of Washington University in St. Louis, which is a member of NNIN and supported by the NSF (Award No. ECS-0335765).